\documentclass{aa}  

\usepackage[]{amsmath}      
\usepackage{graphicx}
\usepackage{color}
\usepackage{txfonts}
\usepackage{url}
\usepackage{natbib}         
\bibpunct{(}{)}{;}{a}{}{,}  
\newcommand{\vc}[1]{\mbox{\boldmath{$#1$}}}  

\begin{document}

   \title{Quasi-static contraction during runaway gas \\ accretion onto giant planets}

   \subtitle{}

   \author{M. Lambrechts
          \inst{1}
          \and
          E. Lega \inst{2}
	  \and
	  R.\,P. Nelson \inst{3}
	  \and
	  A. Crida \inst{2,4}
	  \and
          A. Morbidelli \inst{2}
          }

   \institute{
	   Lund Observatory, Department of Astronomy and Theoretical
	   Physics, Lund University, Box 43, 22100 Lund, Sweden\\
	   \email{michiel@astro.lu.se}
           \and
	   Universit\'e C\^ote d'Azur, Observatoire de la C\^ote d'Azur, CNRS,
           Laboratoire Lagrange, Bd de l'Observatoire, CS 34229, 
           06304 Nice cedex 4, France
	   \and
	   Astronomy Unit, Queen Mary University of London, Mile End Road,
	   London, E1 4NS, U.K.
           \and
           Institut Universitaire de France, 
           103 Boulevard Saint-Michel, 75005 Paris, France
           }

   \date{}

  \abstract{
    Gas-giant planets, like Jupiter and Saturn, acquire massive gaseous
    envelopes during the approximately 3 Myr-long lifetimes of protoplanetary
    discs.
    In the core accretion scenario, the formation of a solid core of around $10$
    Earth masses triggers a phase of rapid gas accretion.
    Previous 3D grid-based hydrodynamical simulations found runaway gas
    accretion rates corresponding to approximately 10 to 100 Jupiter masses per
    Myr.
    Such high accretion rates would result in all planets with
larger-than-10-Earth-mass cores forming Jupiter-like planets, in clear contrast
to the ice giants in the Solar System and the observed exoplanet population.
    In this work, we use 3D hydrodynamical simulations, that include
    radiative transfer, to model the growth of the envelope on planets with
    different masses. 
    We find that gas flows rapidly through the outer part of the envelope, but
this flow does not drive accretion.
    Instead, gas accretion is the result of quasi-static contraction of the
inner envelope, which can be orders of magnitude smaller than the mass flow
through the outer atmosphere.
    For planets smaller than Saturn, we measure moderate gas accretion rates
    that are below 1 Jupiter mass per Myr. 
    Higher mass planets, however, accrete up to 10 times faster and do not
    reveal a self-driven mechanism that can halt gas accretion.
    Therefore, the reason for the final masses of Saturn and Jupiter remains
    difficult to understand, unless their completion coincided with the
    dissipation of the Solar Nebula.
}


   \maketitle
%

\section{Introduction}

Giant planets acquire their gaseous envelopes in a multi-stage process.
When solid bodies grow more massive than the Earth, they start attracting thick
H/He envelopes from the surrounding gas in the protoplantary disc
\citep{Mizuno_1980}. 
In this first phase, the ongoing accretion of solids provides sufficient heat
to support the young atmosphere, which is less massive than the solid core it
surrounds.
However, as the planet grows larger, it can get isolated from the surrounding
planetesimals \citep{Kokubo_1998} and pebbles
\citep{Morby_2012,Lambrechts_2014a,Bitsch_2018}. 
Without solid accretion, compressional heating of the inner envelope becomes
the dominant source of pressure support \citep{LL2017}. 
As a consequence, a second phase is triggered where the envelope of the planet
cools down. 
The luminosity decreases and the planet slowly gains mass
\citep{Bodenheimer_1986, Pollack_1996}.
Interestingly, if the envelope mass grows sufficiently and becomes comparable to
the core mass, this secular envelope cooling sequence can come to an end.
Then, as shown by pioneering work from \citet{Mizuno_1980} and \citet{Stevenson_1982}, the onset of self-gravity triggers a third phase of rapid gas
accretion.
It is this last epoch of atmosphere growth, when envelopes undergo so-called
\emph{runaway gas accretion}, that is the focus of this work.

The process of runaway gas accretion, a term we will loosely use here to
describe gas accretion onto planets that stopped accreting solids and have an
envelope mass comparable to or larger than the core mass, has been modelled in
two different ways.
Initially, 1D hydrostatic time-evolution models were created
\citep{Bodenheimer_1986, Pollack_1996} that are similar to those used for
stellar evolution calculations.
Later, multidimmensional
hydrodynamical models of gas accretion became numerically feasible
\citep{Bryden_1999, Kley_1999, Lubow_1999,Ayliffe_2009}.

Simplified 1D models assume planetary atmospheres that are in hydrostatic
balance at all times \citep{Ikoma_2000, Paps_2005, Mordasini_2012, Piso_2014, Lee_2014, Coleman_2017}. 
By calculating the luminosity, and hence the rate of heat loss, and by assuming
the luminosity is sourced by the accretion of gas onto the planet, it becomes
possible to integrate the model forward in time.
These types of models consistently find that envelopes becoming comparable to
the core mass start to rapidly accrete gas by quasi-static contraction.
After solid accretion has come to a halt, gas accretion first proceeds slowly
and then reaches rates around $10^{-3}$\,M$_{\rm E}$/yr around Saturn-mass
planets \citep{Mordasini_2012}, under nominal conditions.
Nevertheless, the simplifying nature of 1D calculations have made it difficult
to draw firm conclusions. 
Most limiting is the assumption that these planets in 1D models are in perfect
hydrostatic equilibrium all the way out to the edge of the envelope.

Hydrodynamical models in 3D demonstrated that 
hydrostatic balance 
is a problematic assumption.
Generally, protoplanetary discs easily provide gas to the Hill sphere around accreting cores, even when the planet starts carving a gap in the disc \citep{Bryden_1999, Kley_1999, Lubow_1999}.
Therefore, at all times gas can enter the envelope and dynamically interact
with the planet.
However, not all of this inflowing gas becomes bound to the planet.
Indeed, previous studies around low-mass planets, where the envelope mass is smaller than the core mass, have shown that most of the gas that enters the Hill sphere is not accreted, but simply gets advected out of the envelope and redeposited in the disc \citep{Tanigawa_2012,Ormel_2015,Fung_2015,Cimerman_2017,LL2017,Kurokawa_2018,Popovas_2018}. 
Radiative simulations show that only the central envelope, which is shielded from this mass flux, accretes gas \citep{Angelo_2013,Cimerman_2017,LL2017,Kurokawa_2018}.

Around higher mass planets the situation remains unclear 
\citep{Tanigawa_2002, Angelo_2003, Machida_2010,Gressel_2013,Szul_2014}.
These studies report accretion rates much higher than 1D models, of
the order of $10^{-2}$ to $10^{-1}$\,M$_{\rm E}$/yr when planets enter the
runaway regime.
However, these rates were made under approximations to limit the numerical
cost, such as low resolution, the presence of an artificial sink-cell at the
centre of the planet or limitations in the equation of state, such as a
constant temperature approach \citep{Lubow_1999,
Tanigawa_2002,Angelo_2003,Machida_2010,Gressel_2013}.
On the other hand, 3D radiative simulations using SPH reported slower accretion
rates \citep{Ayliffe_2009}. 
However, these rates do not appear to be in agreement with 3D radiative
hydrodynamical simulations by \citet{Szul_2016}.
In summary, the 1D and 3D gas accretion rates reported in the literature
for the runaway gas accretion regime vary widely and the various simplifying assumptions make comparison difficult.

In this work, we measure gas accretion rates onto planets of various masses
ranging from $15$ to $330$\,M$_{\rm E}$, using global 3D simulations that
include radiative transfer. 
A full description of the methods can be found in Section\,\ref{sec:methods}.
By limiting the integration times of our high resolution simulations to tens of orbits, we can measure quasi-steady gas accretion rates without evolving the gravitational potential in time.
Based on these snapshot simulations, we argue that runaway gas accretion
proceeds through quasi-static contraction, as discussed in
Section\,\ref{sec:results}.  Initially, runaway gas accretion is measured to be
relatively slow, below a Jupiter mass per Myr. 
However, planets larger than Saturn accrete at rates that
double their mass in less than $10^5$ yr. 
Then, by combining the sequence of measured accretion rates for given planetary masses, we trace the planetary mass as function of time, 
from the low-mass regime around $10$\,M$_{\rm E}$ up to masses of
fully-formed giant planets larger than $100$\,M$_{\rm E}$
(Section\,\ref{sec:impli}).
In this way, we argue that a planet can grow from approximately $10$\,M$_{\rm E}$ to a giant planet larger than $100$\,M$_{\rm E}$ in less than a $1$\,Myr.
We subsequently discuss the implications of our findings on early and late
formation scenarios for giant planets. 
We summarise our results in Section\,\ref{sec:conc}.

\begin{table}[t!]
  \caption{Simulation parameters.}
  \label{tab:runs}
  \centering
  \begin{tabular}{c c c c c c}
  \hline\hline  
  Name & $M/{\rm M_{\rm E}}$ & $\delta x/d_{\rm H}$ 
  & $r_{\rm min,max}/r_{\rm p}$ & $r_{\rm s}/r_{\rm H}$ & $t$ [$P$] \\
  \hline
  \texttt{run15}    & 15  & 1/40  & 0.7 1.3 & 0.2 & 30 \\
  \texttt{run30}    & 30  & 1/48  & 0.7 1.3 & 0.2 & 30 \\
  \texttt{run100}   & 100 & 1/70  & 0.6 1.4 & 0.2 & 30 \\
  \texttt{run200}   & 200 & 1/84  & 0.6 1.4 & 0.2 & 30 \\
  \texttt{run330}   & 330 & 1/100  & 0.4 1.6 & 0.2 & 25 \\
  \texttt{run100HR} & 100 & 1/230 & 0.6 1.4 & 0.1 & 4 \\
  \texttt{run330HR} & 330 & 1/120 & 0.4 1.6 & 0.1 & 5 \\
  \hline
  \end{tabular}
  \tablefoot{
	  Each simulation is listed in  Col.\,1.  The following columns give
the planetary mass in Earth mass, the resolution (as the inverse of the number
of cells along a diameter in the Hill sphere), the width of the annulus with
respect to orbital radius of the planet, the ratio of smoothing length to the
Hill sphere radius, and the number of orbits performed at highest resolution.
  }
\end{table}

\section{Methods}
\label{sec:methods}

\subsection{Numerical model}
\paragraph{Planet model}
We numerically solve the hydrodynamical equations describing a planet embedded
in an annulus of a protoplanetary disc, together with the equations of
radiative transfer. 
A complete description of our methods can be found in \citet{LL2017} and
a detailed description of the \texttt{FARGOCA} code can be found in \citet{Lega_2014}.

This work differs from previous hydrodynamical works in two ways. 
Firstly, we do not employ a sink cell at the centre of our simulated planet
removing mass or heat.
Secondly, we solve for radiative transfer. This allows us to not be limited to
isothermal numerics, which is important to correctly capture the atmosphere dynamics \citep{Cimerman_2017, LL2017}. 
In order to do so, we make use of the flux-limited diffusion approach \citep[FLD, ][]{Levermore_1981} when solving the energy equation for both the thermal and radiative energy density \citep{Bitsch_2013}. We use the ideal gas equation of state with an adiabatic index of $\gamma=1.4$. 
For the opacity, we employ the prescription provided by \citet{Bell_1994} that covers the opacity provided by the gas and dust component of a gas with
interstellar-medium composition and a solar dust-to-gas ratio of 0.01.

\paragraph{Gravitational potential}
A realistic gravitational potential is hard to obtain for high-mass planets.
Like other studies, we employ a fixed potential for the planet, which does
not take fully into account the self-gravity of an evolving envelope
\citep{Klahr_2006, Kley_2009, Szul_2016}. 
This approach is consistent with our aim of probing the runaway regime through a series of short time integrations around planets of increasing mass.
Additionally, the gravitational potential requires artificial smoothing to
avoid a too strong central mass concentration. 
For the smoothing length we used a constant fraction of the Hill sphere, $r_{\rm
s} = 0.2$\,$r_{\rm H}$ or $r_{\rm
s} = 0.1$\,$r_{\rm H}$ (see Table \,\ref{tab:runs}). 
Here, $r_{\rm H}$ is the radius of the Hill sphere given by
\begin{align}
  r_{\rm H} = \left( \frac{M_{\rm p}}{3M_\odot} \right)^{1/3} r_{\rm p}\,,
  \label{eq:rH}
\end{align}
which corresponds to the maximal gravitational reach of the planet. At larger
radii the tidal gravity force dominates.
A detailed description of the potential can be found in Appendix A of
\citet{LL2017}.

\paragraph{Disc set-up}
We simulate a full annulus of the protoplanetary disc, in $3$ dimensions.
The width of the annulus can be found for each simulation in Table \ref{tab:runs}.
To obtain sufficient resolution in the planetary atmosphere, we made use of a
non-uniform grid \citep{LL2017}. 
Additionally, we make use of mirror symmetry across the midplane to limit our
simulations to the upper hemisphere.
Table \ref{tab:runs} lists the effective resolution for each of our
simulations. A technical discussion of our numerical approach can be found in Appendix\,\ref{ap:num}.

We use nominal values for the disc parameters at the location of the planet at
$5.1$\,AU.
The disc gas surface density is set to $\Sigma_{\rm g}/(M_\odot r_{\rm p}^2) =
6.76 \times 10^{-4} (r/r_{\rm p})^{-1/2}$, corresponding to approximately
$210$\,g/cm$^2$ at the location of the planet. 
We kept the viscosity fixed at $\nu/(r_{\rm p}^2 \Omega_{\rm p}) = 10^{-5}$.
The aspect ratio is maintained by viscous heating. We find an unperturbed
aspect ratio of $H_{\rm p}/r_{\rm p} = 0.04$ at the position where the planet
is inserted. 
We do not include heating by irradiation from the central star. 
This reduces the numerical cost of the simulations without significantly
affecting the gas dynamics in the vicinity of the planet \citep{Lega_2015}.
There is no radial accretion flow towards the star.

\paragraph{Numerical procedure}
In order to trace the growth of a planet across a large mass range, from $15$ to
$330$\,M$_{\rm E}$, 
we measure the accretion rates onto planets of various masses that are obtained from the snapshot simulations listed in Table \ref{tab:runs}.
In appendix \ref{ap:num} we argue these short timescale integrations of tens of orbits are sufficient to measure the quasi-steady accretion rate for a given planetary mass.
The accretion rates from the snapshot simulations are afterwards combined to obtain the planetary mass as function of time through interpolation.

In order to reduce the computational cost, we have introduced two
simplifications.
Firstly, we do not include the self-gravity of the envelope, because the
potential of the planet is held fixed in time. Instead, we opted for a snapshot
approach covering different planetary masses.
In this way, we can use short time integrations to measure the instantaneous
quasi-steady accretion rate (longer integrations than our snapshot runs would have to take
into account the changing potential due to the accretion of gas, see also Appendix \ref{ap:num}).
Additionally, we are resolution limited and therefore not able to include the
mass locked in the deep interior of the planet inside approximately
$0.1$\,$r_{\rm H}$.
Because the deep interior is not modelled, we also avoid modelling its thermal
cooling history, which encompasses a long envelope contraction phase when the
envelope is comparable to the core \citep{Ikoma_2000}.
Instead, we effectively study the contraction of the outer mass layers, which
we argue also for planets in the runaway phase depends on the total mass
potential of the planet.
Below, we describe  in more detail our numerical procedure, which involves
changing both the number of grid cells and the computational domain during the
simulations, so that we arrive at snapshot calculations with the resolution
required to make accurate measurements of the gas accretion rates.

First, before inserting the planet we bring the disc into radiative equilibrium.
This equilibrium is obtained for a 2D $(r,z)$ axisymmetric disc.
The disc annulus extends radially from $r_{\rm min}/r_{\rm p}=0.4$ to
$r_{\rm max}/r_{\rm p}=2.5$. 
Because the planets will be held on fixed non-inclined circular orbits, we can make use of mirror symmetry across the midplane and limit our
simulations to the upper hemisphere.
In the vertical direction the disc extends from the midplane ($\theta =
\pi/2$) to $6^\circ$ above the midplane. 
The resolution is  $(N_r,N_{\theta},N_{\phi})=(224,26,2)$. 
We use periodic boundary conditions in the azimuthal direction.
In the radial direction we use evanescent boundaries to minimize the reflection of density waves \citep{deVal2006}. 
The upper boundary condition is reflective.

We then expand the disc in three dimensions and follow different simulation
strategies for different planetary masses.
For those planets in the mass range from $15$ to $30$\,M$_{\rm E}$, we
simulate a 3D annulus with a restricted radial extent from $r_{\rm min}/r_{\rm
p}=0.7$ to $r_{\rm max}/r_{\rm p}=1.3$.
We then make use of a nonuniform grid in order to obtain sufficient resolution
around the center of the planet, while simulating the full azimuthal range of
the annulus as well \citep[see also][]{LL2017}.
The prescription of the non-uniform grid follows \citet{Fung_2015}, which gives
near-uniform cells inside the Hill sphere of the planet and larger cells farther
out.
We use $(N_r,N_{\theta},N_{\phi})=(200,52,1512)$ grid cells in the radial,
polar and azimuth direction and compute the grid spacing in order to have
respectively $40$ and $48$ grid cells along the diameter of the
Hill sphere for \texttt{run15} and \texttt{run30} (see also
table\,\ref{tab:runs}).
This choice of a nonuniform grid does no-longer allow us to make use of the
large time steps obtained with the FARGO algorithm \citep{Masset_2000}.
We choose a smoothing length of $r_{\rm s} = 0.2\,r_{\rm H}$.
The planetary mass is increased smoothly during $5$ orbits to let the disc 
relax to the presence of the planet.
We run the simulations for a total number of $30$ orbits and measure how
the mass contained within a sphere of radius $0.3$\,$r_{\rm H}$ evolves in time.

More massive planets carve gaps in the disc and therefore we use a different
simulation strategy. 
We now start with expanding the 2D ($r,z$) disc in the azimuthal direction by
using a uniform grid of ($N_r,N_{\theta},N_{\phi})=(224,26,680)$ grid cells. 
By using this uniform grid we benefit from the time-step provided by the FARGO algorithm which allows us to simulate the gap opening process until a stationary state is reached.
At this resolution we have $10$ grid cells in the Hill diameter and therefore
we choose a smoothing length of $r_{\rm s}=0.8\,r_{\rm H}$.
We introduce the planet on a relaxation timescale of $20$ orbits
and run the simulation for an additional $80$ orbits to {\rm approximately}
reach equilibrium after gap opening.
Nevertheless, we acknowledge that reaching a full equilibrium would require us
to extend the simulation for a viscous evolution time of approximately
$10^4$\,orbital periods, as shown in 2D see simulations \citep[][ Appendix A]{Kanagawa_2017}, which is numerically unfeasible for our resolution in 3D. 
Then we double the resolution and half the smoothing length and run the code
for an additional $25$ orbits.
Finally, the code is restarted on a more narrow annulus with a nonuniform grid
chosen to have sufficient resolution in the Hill sphere to measure gas
accretion. 
The width of the annulus contains the gap as well as the damping region and
therefore varies with planetary mass. 
The resolution inside the Hill sphere, smoothing length and radial width of the
annulus are listed in Table\,\ref{tab:runs} for the simulations with $100$,
$200$ and $330$\,M$_{\rm E}$ planets. 
The change of resolution is accompanied by a further decrease of the smoothing
length during one orbit from $r_{\rm s}/r_{\rm H}=0.4$ to $r_{\rm s}/r_{\rm
H}=0.2$. 
At this resolution and smoothing length we run the code for $30$ orbits ($25$
for \texttt{run330}) and measure the gas accretion rate.
The dependence of the accretion rate on resolution and on the smoothing length
is a delicate point. 
Therefore, we investigated further refined simulations (\texttt{run100HR} and
\texttt{run330HR}), which are also discussed in more detail in Appendix\,A.

\begin{figure*}[t]
  \centering
  \includegraphics[width=18cm]{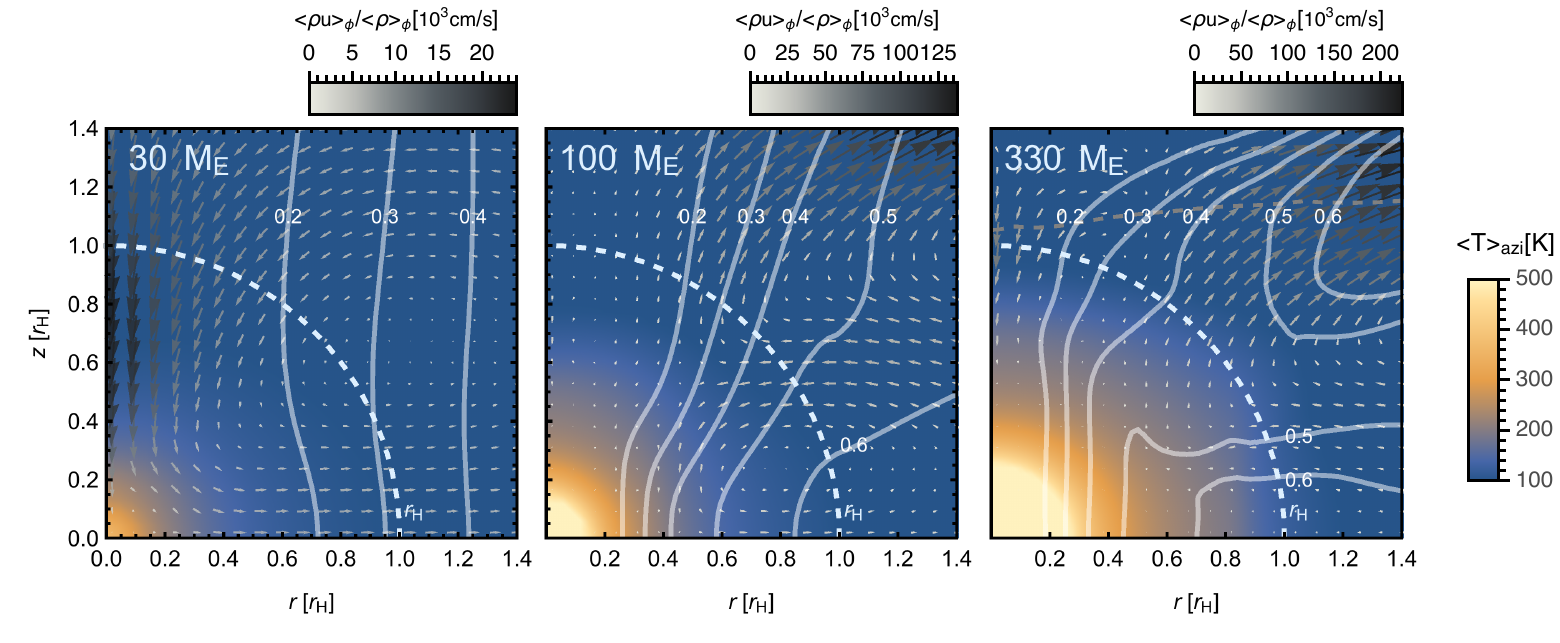}
  \caption{
  The envelope of a growing planet at $3$ different mass stages 
  (left $30$\,M$_{\rm E}$, centre $100$\,M$_{\rm E}$, right $330$\,M$_{\rm E}$).
  The azimuthally-averaged density-weighted mean velocity
  ($\langle\rho\vc u\rangle_\phi/ \langle\rho\rangle_\phi$) 
  in the vertical and radial direction is represented by the
  vector field, 
  using a linear scaling for the magnitude that is also given by
  the grayscale spectrum bar on the top of each panel.  
  The contours show the angular momentum ratio with respect to
  Keplerian rotation, measured in non-rotating co-moving frame.
  The temperature field is color coded in the background.
  In the right panel the horizontal gray dashed line shows where the azimuthally
  averaged optical depth reaches $\tau=2/3$.
  As shown in the left panel, the $30$\,M$_{\rm E}$-planet experiences a high
mass flux from the poles, which moves away from the planet in the midplane
\citep{LL2017}. 
  Higher mass planets also experience a high flux of gas entering
  through the poles, but the process of gap formation changes the
  mass transport and the angular momentum distribution.
  }
  \label{fig:flow}
\end{figure*}

\section{Results}
\label{sec:results}

\subsection{Description of the envelope}

We performed a series of high resolution simulations of planets at different
mass stages.
As the planet grows, the central potential deepens. This causes central densities and temperatures to increase (Fig.\,\ref{fig:flow}).
Only the outer envelope shell experiences rapid gas advection and remains close to the disc temperature.
However, with increasing mass, a larger fraction of the Hill sphere experiences elevated temperatures above $200$\,K.
For the chosen opacity prescription \citep{Bell_1994}, we find that the
envelope within the Hill radius is vertically optically thick, even up to
Jupiter-mass planets, where the optically thin layer is reached at 
$z\approx 1.1\,r_{\rm H}$ (right panel of Fig.\,\ref{fig:flow}).

Initially, low-mass planets below $30$\,M$_{\rm E}$ experience a high mass flux
of gas that enters through the poles and leaves in the disc midplane. Only the
deep interior remains shielded from this flow \citep{Lissauer_2009,Angelo_2013,Cimerman_2017,
LL2017}.
This can be seen from the density-weighted velocity, averaged azimuthally with
respect to the planet, that is shown in the left panel of
Figure\,\ref{fig:flow}.
As the planet carves a gap, the interior within $\approx$\,$0.3$\,$r_{\rm H}$ of planet remains shielded. 
However, a complex flow field emerges, consisting of (i) a continued inwards
flow along the poles of low angular momentum gas with velocities approaching
the sound speed, (ii) a weakly-Keplerian disc and (iii) strong
high-altitude wind flows in the radial direction away from the planet that are
part of large-scale meridional gas flows near the gap edge (right panel of
Fig.\,\ref{fig:flow}).
We further discuss the mass flow and gas accretion in Sec.\,\ref{sec:gasacc},
but we first briefly discuss the angular momentum contained in the envelope.

\subsection{
Rotation in the envelope and implications for satellite formation
}
As the planet grows in mass, the angular momentum distribution inside the envelope changes. 
This evolution can be seen in Fig.\,\ref{fig:flow}, where the white curves show the ratio of the specific angular momentum with respect to the pure Keplerian rotation in the midplane $f= h_z/h_{\rm Kep}$, as measured around the polar axis 
in a frame co-moving with the planet, but not rotating around it\footnote{
I.e., the shown specific angular momentum $l_{\rm inert}$ in the inertial frame
around the planetary polar axis is given by 
$l_{\rm inert} = l_{\rm rot} + \Omega_{\rm K} r^2$, with $\Omega_{\rm K}$ 
the Keplerian frequency of the planet around the sun and $l_{\rm rot}$ 
the specific angular momentum measured in the frame centred on the planet, with one axis along the direction to the star.
}.
Here, $h_{\rm Kep} = \sqrt{GM_{\rm p} r}$ is the angular momentum of gas in Keplerian rotation, expected to be approached when gas settles in a disc around a central potential.
Around low-mass planets ($\lesssim 30$\,M$_{\rm E}$) we find that the angular
momentum distribution is well explained by the circumstellar Keplerian shear penetrating deep
into the 
envelope\footnote{In the non-rotating frame circumstellar Keplerian shear would give
$h_z/h_{\rm Kep} \approx 4^{-1} 3^{-1/2} (r/r_{\rm H})^{3/2}$ \citep{LL2017}.}
\citep{LL2017}. 
Larger planets ($\gtrsim 30$\,M$_{\rm E}$), however, gain prograde angular
momentum. 
However, the angular momentum increase is only moderate, because we do not see
a clear signature of circumplanetary discs in our simulations.
Only Jupiter-mass planets reveal a hint of a disk-like structure which shows strongly sub-Keplerian rotation.
Possibly, this is related to a more 2D-like mass flow around larger mass
planets \citep{Ormel_2015a}.

In the context of the Solar System, the apparent lack of circumplanetary discs, at least around the lower mass planets in the ice giant mass-regime, argues against scenarios where regular satellites form in circumplanetary discs \citep{Canup_2002}.
Therefore, the regular moons around Uranus and Neptune may rather be the product
of late-formed tidal discs of solids that viscously relax and spawn
satellites at the Roche radius \citep{Crida_2012}.

We note that for the $100$\,$M_{\rm E}$ planet the inner $50$\,\% of the Hill
sphere is too warm to allow the condensation of ices, while for the Jupiter
mass planet this region encompasses nearly the whole Hill sphere.  
Such elevated temperatures do not appear to be favorable to the formation of
icy regular moons.
Possibly, this implies that circumplanetary discs only appear late, when the
circumstellar disc cools down and dissipates, which is a topic for further
study.
For now, we can make a crude order of magnitude analysis of the spin angular momentum stored in the bound interior, within $r \lesssim 0.3$\,$r_{\rm H}$. 
We find a centrifugal radius of the bound envelope which is about 
$r_{\rm cfg}/r_{\rm H} \approx 10^{-2} \left(f/0.2 \right)^2 
\left(r/(0.3\,r_{\rm H}) \right)$, 
where the reference value of $f \approx 0.2$ at $r = 0.3$\,$r_{\rm H}$ can be read from Figure\,\ref{fig:flow}.
Here, we assumed no angular momentum loss and ignoring pressure effects and
viscous spreading. 
For comparison, the outermost regular satellite for each giant planet in the
Solar System is situated between 
$10^{-3}\,r_{\rm H}$ (Proteus around Neptune) and 
$5.4 \times 10^{-2}\,r_{\rm H}$ (Iapetus around Saturn).
Further contraction, without angular momentum loss, would imply exceeding break-up velocity. Therefore, it appears that giant planets need to shed their angular momentum efficiently, which may be possible through magnetic coupling between the planet and a circumplanetary disk \citep{Batygin_2018}.

\subsection{Gas accretion}
\label{sec:gasacc}

To a good approximation the interior envelope can be thought of as being in
hydrostatic balance and in the process of slowly contracting, and thus
accreting, over time.
Indeed, the velocity field shown in the different panels of
Figure \ref{fig:flow} reveals that the envelope interior to $\lesssim
0.3\,r_{\rm H}$ is shielded from the mass flux through the outer envelope.

\begin{figure}[t!]
  \centering
  \includegraphics[width=8.8cm]{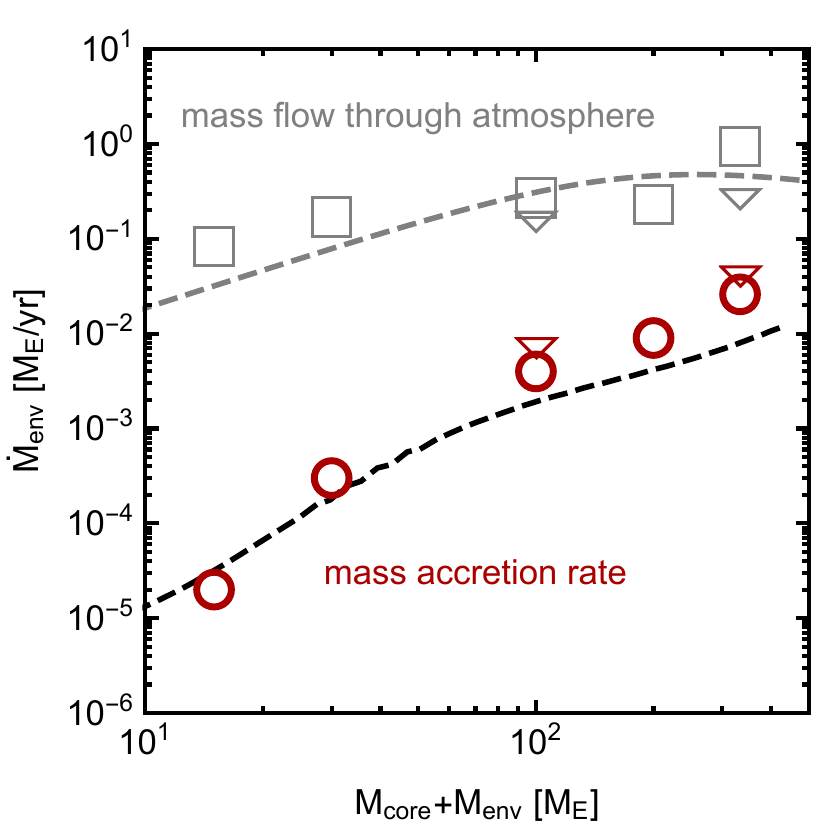}
  \caption{
  Gas accretion rates as function of planetary mass.
  Red circles indicate mass accretion rates obtained from the hydrodynamical
  simulations, for each of the different planetary
  masses we probed.
  Gray squares represent the measured mass flux through the outer envelope,
  measured through the Hill sphere. 
  Gray and red triangles correspond to results from our highest resolution runs
(\texttt{run100HR,run330HR}).
  The black dashed curve shows the evolution of the mass accretion rate
obtained with a simplified 1D model in order to capture the long-term
evolution.
  The gray dashed line shows Eq.\,\ref{eq:massflux}, where we included the
  turn-over around one Jupiter mass as in \citet{Tanigawa_2007}.
  We found this expression corresponds well with the bulk mass flux 
  through the envelope, but not with the gas accretion rate.
  }
  \label{fig:Mdotgas}
\end{figure}

We compute the gas accretion rate by monitoring the increase in the envelope
mass inside $0.3$\,r$_{\rm H}$, during $8$ orbits (see Appendix \ref{ap:num} for
more details).
These measurements are represented by the red symbols in
Figure \ref{fig:Mdotgas}.
In general we find good agreement with similar accretion rate measurements made
using an SPH method \citep{Ayliffe_2009}. However, we do not find a turnover
in accretion rate leading to decreasing accretion rates beyond $100$\,M$_{\rm
E}$. 
As can be seen in Figure \ref{fig:Mdotgas} (triangle symbols), we do find that our accretion rates increase gently when we refine simulations by reducing the smoothing length and increasing the resolution (a more detailed discussion can be found in Appendix \ref{ap:num}).

We verified that these measured accretion rates are consistent with the energy
released by compressional heating in the interior envelope that is dominantly
generated through gas settling in the gravitational potential, through the
relationship $\dot M \approx (GM_{\rm p}/r_{\rm s})^{-1} L$. 
Here $L$ is the luminosity from compressional heating 
(obtained by integrating over the dominant compression term in the energy equation, $-P \cdot \nabla \vc u$, with $P$ the pressure and $\vc u$ the gas velocity)
and $r_{\rm s}$ is the smoothing length. 
This agreement is our first line of evidence for gas being accreted
through quasi-static contraction of the inner envelope.

We also find that the scaling of the accretion rate with planetary mass can be
approximately reproduced with our 1D evolution model (black dashed line in
Fig.\,\ref{fig:Mdotgas}, see appendix \ref{ap:1D} for description of the
model). 
We note here that matching the absolute values of 3D accretion rates with 1D
rates is difficult, because of the various approximations made in the 1D and
3D calculations, which includes the approximate potential description in 3D. 
Therefore, in this work, we do not aim to directly link 1D and 3D models, as done in \citet{Lissauer_2009,Angelo_2013}.
Instead, for now, we simply evolve planets in the toy model with
an effective opacity of $\kappa_{\rm eff}=0.01$\,cm$^2$/g at the radiative
convective boundary. 
This parameter should thus not be thought of as a true opacity, but rather a free parameter that encapsulates our approximations and which effectively allows us to scale the growth rates as $\dot M \propto \kappa_{\rm eff}^{-1}$.
In this way, without aiming for precise quantitative agreement, we do reproduce
the scaling relation of accretion rate with planetary mass.
We find from our 3D simulations that the mass accretion rate scales
approximately with the planetary mass as $\dot M \propto M^{1.5 \--2}$
(Fig.\,\ref{fig:Mdotgas}), in approximate agreement with our 1D toy model.
Therefore, this supports that, to a good approximation, even high-mass planets
accrete through quasi-static contraction and the accreting gas is never in
dynamical free-fall.

In the low-mass regime, below $100$\,M$_{\rm E}$, our 3D accretion rates appear
to be also in crude agreement with previous detailed 1D model studies
\citep{Tajima_1997, Ikoma_2000}.
For example, \citet{Tajima_1997} report accretion rates on the order of
$10^{-4}$\,M$_{\rm E}$/yr around a $30$\,M$_{\rm E}$-planet.
However, at larger planetary masses we see a divergence between our 3D
accretion rates and those found in 1D studies such as those by
\citet{Tajima_1997} and \cite{Ikoma_2000} that report a steeper scaling of the
accretion rate with mass. 
However, a less steep dependency is found by \citet{Mordasini_2012}, as pointed out by \citet{Ida_2018}. 
The exact origin of this difference may be related to the treatment of the
equation of state for the H/He-envelope, in the form of a variable adiabatic
index prescription.
In contrast, in our study, we have used for consistency a fixed adiabatic index ($\gamma=1.4$)
for both the 1D toy model and our 3D simulations.
This choice for the fixed $\gamma$ was motivated by our 3D simulations where,
due to our limited central resolution, even the inner mass layers do not reach
the ionisation/dissociation conditions that require the variable $\gamma$
approach (see also Appendix\,\ref{ap:num}). Further addressing this point would thus require even higher resolution studies.

An important caveat to this study is that we do not yet explore different disc
conditions. 
We expect that lower gas surface densities and lower disc viscosities could
reduce the supply of disc gas to the planet. 
Moreover, we have used here an opacity prescription where, in the
low-temperature regime, opacities are dominated by the dust component. 
In this standard approach, $\mu$m-dust sizes and a dust-to-gas ratio of
$Z=0.01$ are assumed, which is consistent with the interstellar medium
\citep{Bell_1994}. 
In reality, dust growth through sticking may deplete the small $\mu$-sized grains \citep{Brauer_2008}.
The resulting smaller opacities would lead to larger gas accretion rates, as
energy is more easily radiated away.
Although we have attempted to use nominal disc parameters in this study, more
work will be necessary to map out the influence of these parameters (Schulik et
al.\,\emph{submitted}).
Another caveat is that this work does not make use of smoothing lengths below
$r_{\rm s}/r_{\rm H}=0.1$. Therefore we do not fully model the envelope
interior to the smoothing length and this issue is discussed in more detail in
Appendix\,\ref{ap:num}.

\subsubsection{Low-mass planets}
\label{sec:lmplanets}
Planets with masses below $100$\,M$_{\rm E}$ are relatively slow accretors,
with rates as low as $\dot M_{\rm gas }\approx 2 \times 10^{-5}$\,M$_{\rm
E}$/yr around $15$\,M$_{\rm E}$ planets.
Interestingly, they also show the largest difference in the accretion rate
versus the flux of gas that passes through the outer layers of the Hill sphere.
We measure here the mass flux as the mass flux entering the Hill sphere, not
the net flux difference between the mass flow moving in and out
\footnote{
The gas accretion rate is $\dot M_{\rm env} = - \int_V \vc F \cdot \vc n dS$,
with $\vc n$ the unit vector on surface element $dS$ and $V$ the measurement volume, here taken to be a sphere around the approximately bound interior with radius $r=0.3\,r_{\rm H}$.
The mass flux through the envelope we take to be $\dot M_{\rm flux} = - \int_V
(\vc F \cdot \vc n)|_{\vc F \cdot \vc n<0} dS$, with the volume $V$ now given
by the Hill sphere, e.g.\,we only consider the mass flowing into the
volume.} (Fig.\ref{fig:Mdotgas}).
For the $15$\,M$_{\rm E}$ planet, the unsigned mass flux transiting the envelope is of the order of $\dot M_{\rm gas }\approx 10^{-1}$\,M$_{\rm E}$/yr, which is four orders of magnitude larger than the actual accretion rate.

We also find that the mass accretion rate (red symbols) and the transiting mass flux through the envelope (gray symbols) scale differently with respect to planetary mass, as can be seen in Fig.\,\ref{fig:Mdotgas}.
Nevertheless, in the literature this mass flux through the envelope seems in
some cases to have been interpreted as a gas accretion rate
\citep{Angelo_2003,Machida_2010,Tanigawa_2016}.
Indeed, we find that the mass flux through the envelope is well described by the expression
\begin{align}
\dot M_{\rm flux} \approx 
  \,&  0.4 \times
  \left( \frac{M_{\rm p}}{100\,\rm M_{\rm E}} \right)^{4/3}
  \left( \frac{r_{\rm p}}{5\,\rm AU} \right)^{1/2} \times \nonumber\\
& \left( \frac{H_{\rm p}/r_{\rm p}}{0.04} \right)^{-2} 
  \left( \frac{\Sigma_{\rm g}}{410 {\rm g/cm}^2} \right)^{4/3}
  \,\frac{\rm M_{\rm E}}{\rm yr} \,,
  \label{eq:massflux}
\end{align}
which was previously reported as the accretion rate before gap opening \citep{Tanigawa_2002}.
Here, $M_{\rm p}$ and $r_{\rm p}$ are the mass and orbital radius of the
planet. $\Sigma_{\rm g}$ is the gas surface density and $H_{\rm p}/r_{\rm p}$ is the aspect
ratio of the protoplanetary disc, measured at the location of the planet.
The gray dashed line in Fig.\,\ref{fig:Mdotgas} shows Eq.\,\ref{eq:massflux},
but also includes the gap formation induced turn-over around Jupiter mass as in
\citet{Tanigawa_2007} and \citet{Tanigawa_2016}.

Finally, we note that these high mass fluxes through the envelope do not appear
to strongly quench gas accretion rates. This occurs because the cooling
interior of the planet remains shielded from the advection of gas and a
convective-radiative-advective three-layer envelope structure develops
\citep{LL2017}.

\subsubsection{High-mass planets}
\label{sec:hmplanets}

Higher mass planets start carving a gap in the disc, as tidal and pressure torques overcome the viscous torque \citep{Crida_2006}.
However, this process does not halt the flow of gas through the outer atmosphere \citep{Morbidelli_2014}.
We find indeed that the mass flow rates through the envelope remain of the same
order of magnitude and are comparable to the flux given in
Eq.\,(\ref{eq:massflux}). 
At most, the decrease in the local gas surface density due to gap formation leads to the rate of mass flux through the envelope leveling off around planets larger than $200$\,M$_{\rm E}$ \citep{Tanigawa_2007,Tanigawa_2016}.

Gas accretion rates similarly appear not affected by gap formation.
The accretion rates continue to grow with mass, as can be seen in Fig.\,\ref{fig:Mdotgas}, and follow the same trend as our 1D-model.
Gas accretion rates become as high as $10^{-2}$\,M$_{\rm E}$/yr around Jupiter mass planets.
We nevertheless caution that further work is required to study gas accretion in
the context of gap formation in low viscosity discs with long viscous
equilibration timescales \citep{Kanagawa_2017}.

Can a disc supply these high gas flow and gas accretion rates?
Our simulations cannot address this question directly as they only cover an annulus of a gas disc during a short timescale.
That being said, gas accretion rates onto solar-like stars can initially be as
high as approximately $10^{-7}$\,M$_\odot$/yr, while decaying to about
$10^{-9}$\,M$_\odot$/yr near the end of the disc lifetime
\citep{Antoniucci_2014,Manara_2016,Hartman_2016}. 
This would correspond to a global flow rate of gas through the disc
of approximately $3.3\times 10^{-2}$ to $3.3\times 10^{-4}$\,M$_{\rm E}$/yr. 
Therefore, in the first few Myr, discs can provide gas to accreting giant planets.
Only gas giants that emerge near the end of the disc lifetime may see
diminished gas accretion rates. 
However, even these giant planets would still be able to double their mass in
about a Myr.
That being said, the complicated interplay between gas accretion, gap formation
and disc depletion warrants further study.

\section{Implications}
\label{sec:impli}

Observations of short-period giant exoplanets, that are larger than
$15$\,M$_{\rm E}$, show a wide and continuous range of envelope-to-core mass
ratios, ranging from planets where the envelope mass barely exceeds the core
mass, up to gas giants as large as about $4$\,Jupiter masses\footnote{
The occurrence rates of larger stellar companions do not longer increase with stellar metallicity and therefore likely no longer trace planets formed in the core accretion scenario \citep{Santos_2017}.
}.
This is surprising given that planets larger than $15$\,M$_{\rm E}$ are
susceptible to runaway gas accretion, if they form sufficiently early to spend
more than approximately a Myr in the gas disc \citep{LL2017}. 
Runaway gas accretion would only produce planets with high
envelope-to-core mass ratios, contrary to these observations.
Why some cores accrete thick envelopes, and others do not, is difficult to
explain within our current theoretical understanding. We consider two cases, late and early formation.

\begin{figure}[t!]
  \centering
  \includegraphics[width=8.8cm]{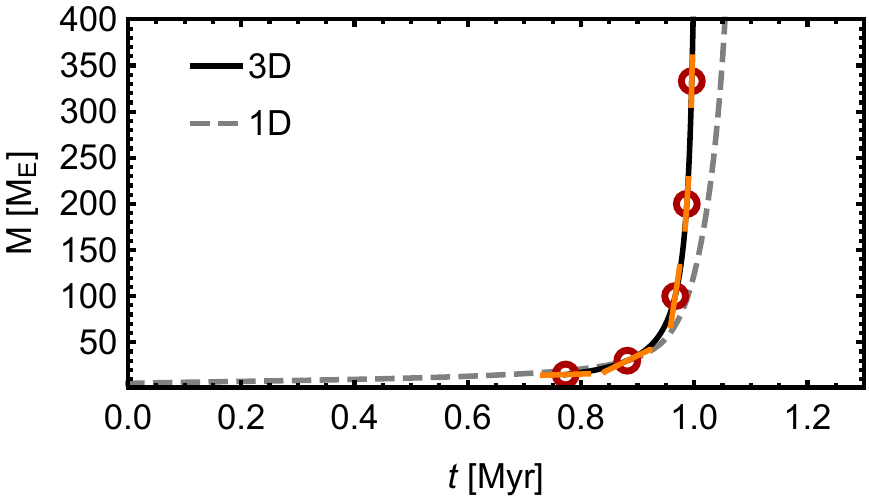}
  \caption{
  Evolution of the planetary mass as function of time, based on the snapshot
accretion rates of our hydrodynamical simulations (black curve). 
  Red circles mark the masses where hydrodynamical simulations were performed. 
  The measured accretion rate of each simulation is indicated by the orange
slope segment.
  Growth from a $15$\,M$_{\rm E}$-planet to a Jupiter-mass planet is completed
in about $2\times 10^5$\,yr.
  The gray dashed curve represents the 1D model, illustrating a significant fraction of envelope growth may be spend at low envelope masses.
  }
  \label{fig:Menv}
\end{figure}

Our study supports a scenario where giant planets reach the point of runaway
gas accretion late in the approximately 3 Myr-lifetime of the protoplanetary
disc.
We find increasing gas accretion rates for planets of increasing
mass, such that when they are joined in an evolutionary sequence they
correspond to Jupiter growing to completion in $1$\,Myr, as illustrated in Figure \ref{fig:Menv}.
Here, the black curve shows the time evolution of the runaway gas phase, obtained by integrating a least-squares fit of a power law to the measured accretion rates of our hydrodynamical simulations. We find $\dot M_{\rm env} \approx 2.7 \times 10^{-3} \left( M/(100\,{\rm M}_{\rm E}) \right)^{1.9}$.
We prefer this over piece-wise integration due to the sparse mass sampling. 
Given the previously discussed uncertainties in the accretion rates, the time obtained here for runaway gas accretion is only approximate. 
We find that the growth from $15$ to $330$\,M$_{\rm E}$ would be completed in about $2\times 10^5$\,yr. 
However, most of the time growing is spent when the planet is smaller than
$15$\,M$_{\rm E}$, as indicated by the 1D model (gray dashed curve). 
Therefore, mainly for illustrative purposes, we have shifted the black growth
curve by $\approx 0.8$\,Myr to match the point where the 1D planet reached
$30$\,M$_{\rm E}$, but note that this early contraction phase is sensitive to
when solid accretion halted onto the planet and the final core mass
\citep{LL2017}.
That being said, the short time-scale for runaway gas accretion we measure here
would thus argue for scenario where giant planets like Jupiter emerged late
in the disc lifetime, in order to explain their present day mass.

This late-formation view is possibly supported by the emerging populations of
exoplanets detected through microlensing surveys \citep{Suzuki_2016}.
This technique probes exoplanets down to approximately Neptune mass, in orbits
around the ice line of their host star, which is somewhat similar to the giant
planets we consider in this work.
\citet{Suzuki_2016} report a single power law distribution for planet occurrence with respect to the planet-to-host-star mass $q$. 
They find that $d N/d \log q \propto q^n$, with $n = - 0.93 \pm 0.1$ holds well
between Neptune to 10-Jupiter-mass planets.
This also appears consistent with a recent analysis of the long-period
transiting planet sample from Kepler \citep{Herman_2019}.
Thus, at least for this population of relatively wide-orbit planets,
there appears to be no signature of a natural mass where accretion comes to a
halt 
\citep[which would translate in a sharp increase in the number of planets at
that mass,][]{Suzuki_2018}.
Instead, the steep slope in occurrence rates seems more consistent with
continued mass growth up until gas is removed from the protopanetary disc (a process independent of the planet mass).
In fact, the slope of the mass distribution is a measure of the scaling of the
gas accretion rate with mass, assuming a steady state distribution. 
Because $dN/dq \propto (dN/dt) (dq/dt)^{-1}$, it appears that giant planets grow
approximately along $\dot M \propto M^{-n+1} \propto M^{1.9}$. 
Encouragingly, the observed mass accretion scaling appears to be consistent
with the mass scaling we find numerically where 
$\dot M \propto M^{1.5 \--2}$ (Fig.\,\ref{fig:Mdotgas}) and 
a least-squares fit gives $\dot M \propto M^{1.9}$.  

Alternatively, one can hypothesise runaway gas accretion occurred early in the
evolution of the disc. 
Planetary cores could emerge within a Myr timescale in a pebble accretion scenario \citep{Lambrechts_2012,Lambrechts_2014b,Bitsch_2015}.
Such early core formation has been proposed for Jupiter in order to separate
the inner and outer Solar System into two distinct isotopic reservoirs after
only about 1\,Myr of disc evolution \citep{Kruijer_2017}. 
However, the fast appearance of giant planets is problematic for two reasons. 
Firstly, given the accretion rates we report, one needs to invoke an as
of yet unknown physical mechanism to limit gas accretion onto the planet 
for potentially several Myr before the disc dissipates.
Secondly, even if accretion could be halted, early-formed gas giants
would migrate rapidly through discs \citep{Bitsch_2015}, because of type-2
migration (\citealt{Lin_1986}, but see also \citealt{Durmann_2017}, \citealt{Kanagawa_2018} and \citealt{Robert_2018}).

In summary, this work argues gas giant reached the point of runaway growth late in the disc lifetime, possibly due to slow pre-runaway gas accretion or the relatively late formation of planetary cores. 
However, more work will be necessary to understand the physical mechanisms
behind the final masses, and orbital locations, of giant planets.
A promising avenue for future work could be the study of planets in stratified
low-viscosity disks where accretion and migration may be slower.

\section{Conclusions}
\label{sec:conc}
In this paper, we have numerically investigated how giant planets accrete
their gaseous envelopes. 
For planets between  15\,M$_{\rm E}$ and 1\,M$_{\rm J}$ in mass, 
we have measured that gas accretion proceeds through quasi-static contraction
of a nearly hydrostatic envelope that is located within the inner $\sim 30$\,\%
of the Hill radius.

The accretion rate of material onto the inner bounded envelope is between $2$
to $4$ orders of magnitude lower than mass flux of gas into the Hill sphere. 
Moreover, the effective accretion rate shows a different scaling relation with
planetary mass compared to the mass flux through the Hill sphere.
Therefore, the complex 3D gas flow in the outer envelope is limited to the
advection of gas in and out of the Hill sphere, and unrelated to the gas
accretion rate onto the planet.
We do not identify the presence of a circumplanetary disc around accreting planets smaller than about $100$\,M$_{\rm E}$ in mass.

Our model suggests that, after the emergence of an approximately $15$\,M$_{\rm
E}$-planet, the formation of a Jupiter-mass planet can occur within approximately $2\times10^5$\,yr at 5\,AU, assuming an ISM-like opacity.
These growth rates are however dependent on the opacity, viscosity and gas
surface density, and these dependencies need further exploration.
Moreover, our study makes use of a smoothing parameter that regulates how
centrally condensed the gravitational potential is. 
Therefore our results are focussed on the description of the envelope outside
of approximately $0.1$\,r$_{\rm H}$.
Further high-resolution 3D simulations with radiative transfer, that include
high-temperature changes in the equation of state, and ideally the inclusion of
self-gravity, will be needed to explore the role of the deep envelope interior
within $0.1$\,r$_{\rm H}$.
An interesting prospect is that such work would allow probing the convective interior and characterise the radiative-convective boundary, an important transition that would facilitate the comparison to 1D models \citep{Ikoma_2000, Piso_2014}.

Finally, our radiative hydrodynamical simulations do not reveal any thermally
induced processes that can strongly reduce the accretion of massive gas
envelopes on Myr timescales. 
In the absence of other processes that can slow down accretion, this implies
that the masses of the giant planets in the Solar system are most naturally
explained because they formed in a limited gas reservoir, most likely due to
forming late in the lifetime of the protoplanetary disc.

\begin{acknowledgements}
M.L.\,likes to thank 
  Bertram Bitsch, 
  Matth\"aus Schulik,
  Anders Johansen and 
  Yuri Fujii for stimulating discussions.
  The authors are grateful for the constructive feedback by an anonymous
  referee. 
M.L., E.L., A.C. and A.M.\,are thankful to ANR for supporting the MOJO project (ANR-13-BS05-0003-01).
R.P.N.\,acknowledges support from the STFC grants ST\/P000592\/1 and
ST\/M001202\/1.
This work was performed using HPC resources from GENCI [IDRIS] (Grant 2017/2018,
[i2016047233]) and Mesocentre SIGAMM, hosted by the Observatoire de la
C\^ote d'Azur.  
\end{acknowledgements}

\bibliographystyle{aa}        
\bibliography{references}     

\appendix

\section{Numerics}
\label{ap:num}

\begin{figure}[t!]
  \centering
  \includegraphics[width=8.8cm]{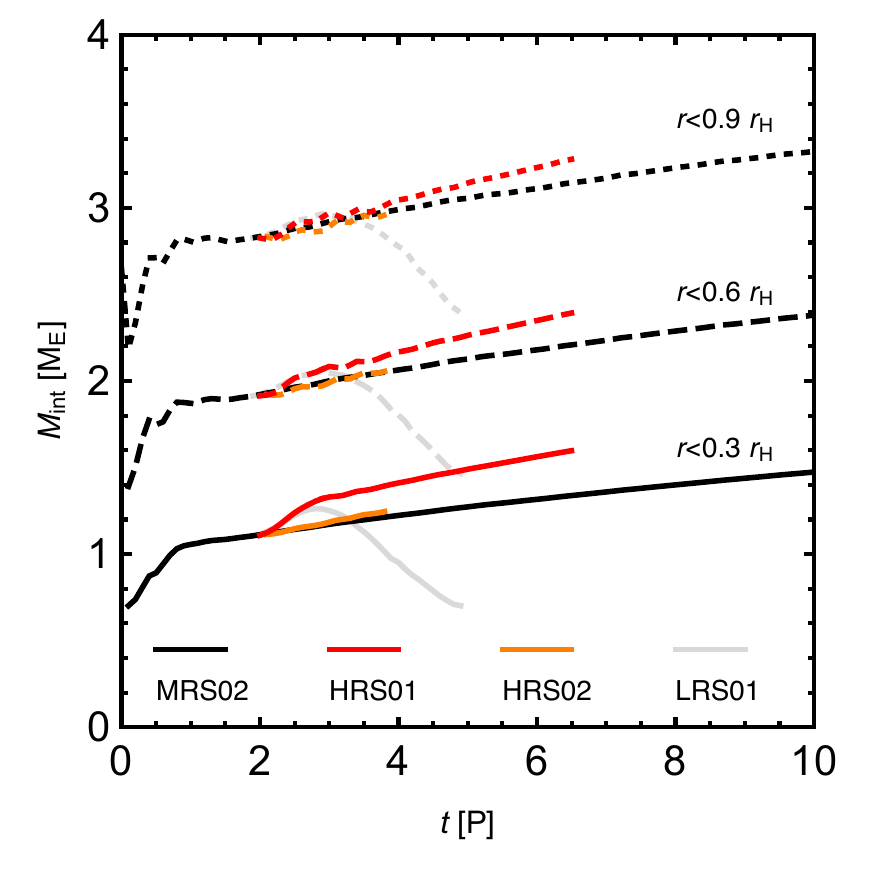}
  \caption{
  Envelope shell mass as function of time, interior to $0.3,0.6,0.9$\,$r_{\rm
H}$.  for the Saturn-mass case.
  Black curves correspond to \texttt{run100}, labeled as MRS02 (medium resolution, smoothing length $r_{\rm s}/r_{\rm H} = 0.2$). 
  The smoothing length is relaxed to its final value over one orbit,
  corresponding to the initial increase in envelope mass.
  Further gas accretion is the result of the physical cooling of the envelope. 
  Increasing the resolution even further does not change the accretion rate, as
indicated by the orange curve (labeled \texttt{HRS02}).
  A refined simulation with twice the resolution and half the smoothing length
($r_{\rm s}/r_{\rm H} = 0.1$, \texttt{run100HR}) is shown in red. 
  The gray curves represent a test simulation with factor 2 smaller smoothing
length, but without the increase in resolution, which results in an unstable
envelope. 
  }
  \label{fig:convergence}
\end{figure}

\paragraph{Resolution}
We have tested our results against changes in resolution and the choice of the smoothing length of the potential.
This is illustrated in Fig.\,\ref{fig:convergence}. 
Here, the mass in an envelope shell is plotted as function of time for
\texttt{run100} (black curves). 
After relaxing the potential over an orbital period, the interior envelope, within $r \lesssim 0.3 r_{\rm H}$, starts to contract and accrete mass. 
We find then, similar to \citet{LL2017}, that the luminosity of the planet reaches an approximate steady state balance after the adjustment the changing potential.
As shown in Fig.\,\ref{fig:convergence}, the outer envelope shells do not participate significantly in the accumulation of mass. Because mass growth occurs in the centre, it is necessary to model the interior with sufficient resolution.

For a fixed smoothing length, we empirically identified that when using above
$8$ cells per smoothing length we obtain convergent behaviour in the interior
structure and accretion rate. 
Indeed, a simulation with twice the resolution gives similar accretion rates (orange curves, labeled \texttt{HRS02}).
Conversely, in simulations where a too low resolution per smoothing length is
introduced, the balance between gravity and pressure support can get destabilized.
This is illustrated with a test run indicated by the gray curves (labeled \texttt{LRS01}) in Figure\,\ref{fig:convergence}. 
Therefore, our standard practice when halving the smoothing length is to double the resolution, we term this procedure refinement.

As mentioned in Sect.\,\ref{sec:gasacc}, accretion rates increase moderately when we refine the simulations. This can also be seen in Fig.\,\ref{fig:convergence}, where \texttt{run100} (black curve) can be compared with \texttt{run100HR} (red curve).
We thus find slightly higher accretion rates with increasing refinement.
This physical effect occurs because reducing the smoothing length
effectively opens up a new interior region of the planet. 
A refinement by a factor of $2$ between \texttt{run100} and \texttt{run100HR} resulted in an increase of the mass accretion rate by a factor of about $1.5$. 
Therefore, further reductions of the smoothing length could increase the
accretion rate more, but high resolution studies indicate that little envelope
contraction occurs in the strongly pressure-supported central envelope inside
of $0.1$\,r$_{\rm H}$ \citep{Szul_2016}. 
This is also supported by the approximate agreement in measured accretion
rates for $100$\,M$_{\rm E}$-planets between our work and the SPH results of
\cite{Ayliffe_2009} that use $ r_{\rm s}/r_{\rm H} < 0.01$.
Nevertheless, future studies aimed at resolving the deep interior will
also need to address temperature dependent changes of the equations of state
\citep[further discussed below, see also][]{Szul_2016} and should strive for a
more realistic description of the gravitational potential by including
selfgravity.
Taken together, this does indicate that our study is limited to characterising
accretion of envelope down to $\approx 0.1$\,r$_{\rm H}$ and further work is
required to reveal the contraction of the deep interior, but for now, such
higher resolution studies with $ r_{\rm s}/r_{\rm H} < 0.1$ are currently
numerically unfeasible for our model setup.
We further note that deepening the gravitational potential, in principle down
to the core surface, would also require longer envelope equilibration times
exceeding 10 orbital timescales (Fig.\ref{fig:convergence}), which would
further increase the numerical cost.

In our simulations we note a tendency for the accretion rate to start to decrease on long timescales ($\gtrsim 10\,P$). 
This is an artefact of our snapshot approach, where we do not take into account the effect of the accreted mass on the gravitational potential, 
which already starts to exceeds the $1\%$ of the potential mass after 10 orbits
for our Saturn-mass case.
As a result, we pragmatically constrain our measurements of the accretion rates to be made within $10$ orbits.

\paragraph{Equation of state}
We verified that in our current simulations temperatures and densities are such
that our ideal gas equation of state is not violated.
For our Jupiter-mass simulation, temperatures in the most interior shell reach $T=1500$\,K and densities of $\rho = 7.7 \times 10^{-9}$\,g/cm$^3$. 
Therefore we do not yet reach the conditions where H/He ionisation starts playing a significant role \citep{Piso_2015, Popovas_2016}. 
However, future work making use of even more reduced smoothing lengths will
probe more interior regions and will thus also require a more complete equation
of state.

\section{1D toy model}
\label{ap:1D}
We make use of a simple 1D toy model to calculate the mass growth of an
envelope.
It assumes that the growth in atmospheric mass is the result of the competition
between the gravitational contraction of the envelope and how
efficiently the planet transports this heat release.
In practice, we construct a cooling sequence. Subsequent stages of the envelope increase in mass, but see their total energy budget decrease due to release of heat, corresponding to a luminosity:
\begin{align}
  L \approx -\frac{d(E_{\rm th}+E_{\rm grav})}{dt}
\end{align}
where $E_{\rm th}$ and $E_{\rm grav}$ are, respectively, the integrated
thermal and gravitational energy stored in the envelope.

In order to time evolve the envelope, we perform an iterative procedure,
where we require the energy loss during a timestep to be consistent with energy difference between subsequent envelope structures, which are assumed to be in hydrostatic equilibrium. 
This is a common technique used in 1D-models 
\citep{Bodenheimer_1986, Pollack_1996,Ikoma_2000, Paps_2005, Mordasini_2012, Piso_2014, Lee_2014, Coleman_2017}.  
In practice, we first calculate the envelope structure for a given envelope
mass. 
We then time step and predict the subsequent envelope mass. 
The chosen envelope mass fixes the envelope structure and the energy
budget. 
We then perform a convergence step, where we vary the envelope mass until we
reach energy balance.
An in detail prescription of this numerical approach will be given in
Lambrechts and Johansen \emph{in prep}.

This procedure is based on several simplifying assumptions, in order to efficiently calculate the long-term evolution of the planet. 
We assume spherical symmetry. 
Moreover, we assume that envelopes are in perfect hydrostatic equilibrium, from
the core to the outer boundary, here set to be the Hill sphere.
At this outer boundary the envelope connects to the unperturbed nebular temperature $T$ and density $\rho$. 
The outer radiative region is assumed to be isothermal. 
Additionally, the opacity at the radiative convective boundary is held fixed over time. 
Finally, the inner convective interior is approximated as a self-gravitating
polytrope with adiabatic index $\gamma=1.4$, which is solved for through the
Lane-–Emden equation. 
In conclusion, in the toy model we assume that planets evolve at all times,
and for all masses, through quasi-static contraction. This allows us to use the total energy budget of the planet to evolve the planet forward in time, without the need for numerically expensive radiative transfer,  but at the cost of a priori difficult to motivate assumptions on the structure and evolution of the envelope.

\end{document}